\documentclass[12pt]{article}

\usepackage{rotating}
\usepackage{longtable}
\usepackage{threeparttablex}

\usepackage{scicite}

\usepackage{times}
\usepackage[normalem]{ulem}
\usepackage{graphicx}
\usepackage{amssymb}
\usepackage{amsmath}
\usepackage{graphicx}
\usepackage{dcolumn}
\usepackage{bm}
\usepackage{siunitx}
\usepackage{graphicx}
\usepackage{xcolor}
\usepackage{xspace}
\usepackage{multirow}
\usepackage{tablefootnote}
\usepackage{CJKutf8}            

\usepackage{hyperref}

\makeatletter

\makeatletter

\newenvironment{sciabstract}{%
\begin{quote} \bf}
{\end{quote}}

\title{A MUSE Source-Blind Survey for Emission from the Circumgalactic Medium}

\author{Huanian Zhang \begin{CJK*}{UTF8}{gkai} (张华年) \end{CJK*}$^{1,2 \ast}$ \& Dennis Zaritsky$^2$ \\ 
\\ 
\normalsize{$^{1}$Department of Astronomy, 
Huazhong University of Science and Technology, Wuhan, Hubei 430074, China} \\
\normalsize{$^{2}$Steward Observatory, University of Arizona, Tucson, AZ, USA} \\ 
\\
\normalsize{$^\ast$ E-mail: huanian@hust.edu.cn}
}

\date{}

\begin{document}

\baselineskip24pt

\maketitle

\begin{sciabstract}
The recent detection of optical emission lines from the circumgalactic medium (CGM) \cite{Tumlinson2013,Werk2014,CGM2017} 
in combined, large samples of low-redshift, normal galaxy spectra\cite{Zhang2016, zhang2018a, Zhang2018b, Zhang2019, Zhang2020a, Zhang2020b, Zhang2021, Zhang2022} hints at the potential to map the cool ($\sim$ 10$^4$ K) CGM in individual, representative galaxies.
Using archival data from a forefront instrument (MUSE) on the VLT, we present a source-blind, wide-redshift-range ($z \sim 0-5)$ narrow-band imaging survey for CGM emission.  
Our detected, resolved emission line sources are cataloged and include a 30 kpc wide H$\alpha$ source likely tracing the CGM of a  low-mass galaxy (stellar mass $\sim$ $10^{8.78\pm0.42}$ M$_\odot$)\cite{Contini2016} 
at $z=0.1723$,  a 60 kpc wide Ly $\alpha$  structure associated with a galaxy at $z=3.9076$, and a 130 kpc ($\sim r_{\rm vir}$) wide [O II] feature revealing an interaction between a galaxy pair at $z=1.2480$. The H$\alpha$ velocity field for the low-mass galaxy suggests that the CGM is more chaotic or turbulent than the galaxy disk, while that for the interacting galaxies shows large-scale ($\sim 50$ kpc) coherent motions. 
\end{sciabstract}

\section{Introduction}

The circumgalactic medium (CGM) of galaxies contains the majority of a galaxy's baryons\cite{Behroozi2010, McGaugh2010, CGM2017} and has been primarily studied along a limited number of sightlines where the CGM imprints itself onto the spectra of bright background sources\cite{Tumlinson2013,Werk2014,CGM2017}. The recent detection of optical emission lines\cite{Zhang2016, zhang2018a, Zhang2018b, Zhang2019, Zhang2020a, Zhang2020b, Zhang2021, Zhang2022} motivates and guides our study of archival integral field spectroscopy. The first detection of  H$\alpha$ and [N{ \small II}] $\lambda$6583 emission that extends to projected radii of $\sim$ 100 kpc from low-redshift, normal galaxies\cite{Zhang2016} used massive stacks of Sloan Digital Sky Survey (SDSS)\cite{sdss, SDSSDR16} spectra. Subsequent studies characterized the line-emitting, cool CGM within 50 kpc or 0.25 $r_{\rm vir}$ in low redshift, normal galaxies\cite{zhang2018a, Zhang2018b, Zhang2019, Zhang2020a, Zhang2020b, Zhang2021, Zhang2022}. These studies inform expectations of the emission line fluxes for `average' galaxies as a function of radius, stellar mass, environment,  orientation angle, etc. Noting that an 8-m telescope will have a sensitivity that is about one order of magnitude greater than the SDSS for the equivalent exposure time and on-sky aperture size and utilizing binning in both radial and azimuthal directions (total $\sim$ 30 bins), we estimate that 8-m class telescopes can reach the required sensitivity level to detect the CGM in individual, nearby, normal galaxies with total exposure times of one to several nights. Given these lengthy exposure times, gains in efficiency are paramount. 

As such,  integral field spectroscopy on an 8-m class telescope provides the opportunity to simultaneously search for emission over a wide range in redshift and across many galaxies. The Multi Unit Spectroscopic Explorer (MUSE)\cite{MUSE2010}, an integral field unit (IFU) mounted on the Very Large Telescope (VLT-{\it Yepun}, UT4), has a field of view (FOV) of 1  arcmin$^2$ and is well suited for this purpose  because of its comparably large FOV and wide wavelength coverage. Because of the speculative nature of this program and the large required investment of observing time, using archival public data proved to be the simplest way to build a demonstration case for CGM mapping of  individual galaxies. Using MUSE data, which covers wavelengths from 4750 \AA \ to 9300 \AA, we can simultaneously construct images of  Ly $\alpha$, [O {\small II}], [O {\small III}], H$\alpha$ and [N {\small II}] emission originating from the cool CGM gas in targets across a range of redshifts (Table \ref{tab:range}).

\begin{table}[ht]
    \centering
    \caption{{\bf Accessible Emission Lines and Redshifts}}
    \vspace{0.25cm}
    \begin{tabular}{cc}
    \hline \hline
    Emission Line & Redshift Range \\
      \hline
      Ly $\alpha$ &  $3.1 < z < 5.7$ \\ \\
$[$O III$]$ $\lambda \lambda$3727,3729 & $0.3 < z < 1.29$  \\ \\ 
$[$O III$]$ $\lambda$5007 & $z < 0.85$  \\ \\ 
H$\alpha$ & $z < 0.2$  \\  \\
$[$N II$]$ $\lambda$6583 & $z < 0.2$ \\
\hline \hline
    \end{tabular}
    \label{tab:range}
\end{table}

A publicly available dataset that is suitable for this study exists. 
Bacon et al. (2015)\cite{MUSE-HDFS} obtained deep MUSE observations of the {\it Hubble} Deep Field South (HDFS) using approximately 27 hours of exposure time. 
We present an image of the MUSE HDFS field obtained by summing the flux over the entire MUSE optical wavelength range (4750 \AA \ to 9300 \AA) from the data-cube presented by Bacon et al. (2015) \cite{MUSE-HDFS} in Figure \ref{fig:opticalMap}.
In the analysis of these data that we present below, we adopt a $\Lambda$CDM cosmology with parameters $\Omega_m$ = 0.3, $\Omega_\Lambda =$ 0.7, $\Omega_k$ = 0, and $h = $ 0.7\cite{riess,Planck2018}.

We do a source-blind survey for emission by sliding an 8-pixel (10 \AA) filter window, constructing effectively narrow-band images, and identifying significant ($>$5$\sigma$) detections using
SEP\cite{sep}, which is the python version of SExtractor\cite{SExtractor}, and match, when possible, emission line sources we identify to sources in the Bacon et al. (2015)\cite{MUSE-HDFS} and Contini et al. (2016)\cite{Contini2016} catalogs
(see Methods section for more details). The corresponding velocity resolution of this procedure varies from 300 km s$^{-1}$ to 175 km s$^{-1}$ for the different emission structures, respectively, which roughly matches the internal kinematic width of galaxies across a wide range of stellar masses.  We identify a total of 47 emission line sources, 43 of which are associated with previously identified continuum-detected sources and have measured redshifts that are included in Table S1 in the Supplementary Materials. The four remaining objects are matched to continuum sources in the Bacon et al. (2015) catalog, but do not have published redshifts. We present redshift measurements for two that are matched to sources with notations of emission lines in the Bacon et al. catalog (a Ly $\alpha$ source with an M606-band magnitude of 28.49 [AB mag]) and an [O {\small II}] source with an M606-band magnitude is 25.19 [AB mag]). For a third, located at $\alpha = 22 {\rm h} 32^\prime 56.45^{\prime \prime}$, $\delta = -60^\circ 33^\prime 32^{\prime \prime}$ ($\alpha = 338.23523$, $\delta = -60.55887$), we attribute the emission to Ly$\alpha$ at $z = 5.6851$ after examination of narrow-band imaging and the spectral line shape. The remaining  object is a single-line detection and near the edge of the survey region, where the noise is larger. We do not assign a redshift to this source.

\section{Results}

Among the catalogued sources, there are five where we measure that the ``radius" (see Methods section) of the emission line detection is $>$ 4 arcsec. From among these, we select to highlight the sources with the largest angular extent in either the Ly $\alpha$, [O II], or H$\alpha$ lines, each line selects for objects at a different redshift range, and discuss implications for CGM mapping of individual galaxies:

\subsection{A low redshift, low mass galaxy with a tentative CGM detection. } 
\label{sec:low_mass}

The original HDFS target field selection avoided luminous, nearby galaxies for obvious reasons, which limits the list of suitable low-z targets in the field for our purpose. In previous papers presenting the SDSS stacking analysis\cite{Zhang2016, zhang2018a,Zhang2018b, Zhang2019, Zhang2020a, Zhang2020b, Zhang2021, Zhang2022}, we selected galaxies by redshift ($0.02 < z < 0.2$),  half light radius ($1.5 < R_{50}/{\rm kpc} < 10$),  and  luminosity ($10^{9.5} < L /L_\odot < 10^{11}$). There is only one such galaxy within the MUSE HDFS survey that more or less satisfies these same criteria  can serve as a test case for deep IFU observations of nearby galaxies. It is located at $\alpha = 22 {\rm h} 32^\prime 54.68^{\prime \prime}$ and $\delta = -60^\circ 33^\prime 33^{\prime \prime}$ ($\alpha = 338.22784$, $\delta = -60.55921$). Its redshift is 0.1723 and its M606-band magnitude is 21.71 [AB mag]\cite{MUSE-HDFS}.
The extent of the optical disk of this galaxy, where the disk size represents the scale length of an exponential model of the disk measured from Hubble Space Telescope ({\sl HST}) images is $\le$ 2 arcsec \cite{Contini2016}, corresponding to $\lesssim$ 6 kpc at the redshift of the target galaxy.  The stellar mass of the target galaxy is estimated to be $10^{8.78\pm0.42}$ M$_\odot$  and the star formation rate (SFR)\cite{Contini2016} is $0.10^{+0.36}_{-0.08}$ M$_\odot$ yr$^{-1}$. The stellar mass of our target galaxy  is slightly less than that of the Large Magellanic Cloud (LMC), between 2 to $3 \times 10^{9}$  M$_\odot$ within 4 to 9 kpc radii\cite{vanderMarel2002,Eskew2012}. The SFR of the target galaxy is also similar to that of the LMC, which is estimated to be $\sim 0.2$ M$_\odot$ yr$^{-1}$\cite{Harris2009}. 
The virial radius of the target galaxy is estimated to be $\sim$ 67 kpc at the redshift of the target galaxy from the scaling relation between stellar mass and virial radius derived from UniverseMachine\cite{Behroozi2019}. 

\begin{figure}[ht]
\begin{center}
\includegraphics[width = 0.32 \textwidth]{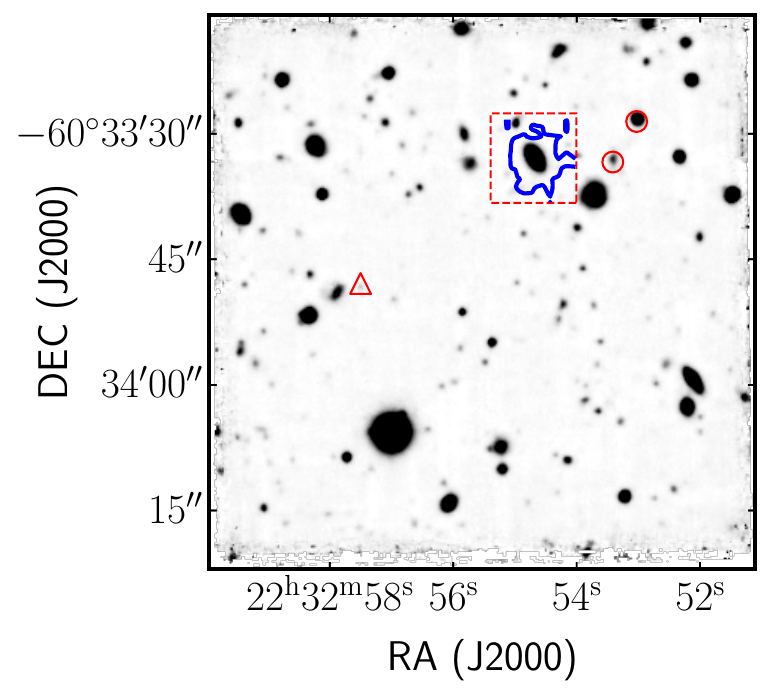}
\includegraphics[width = 0.32 \textwidth]{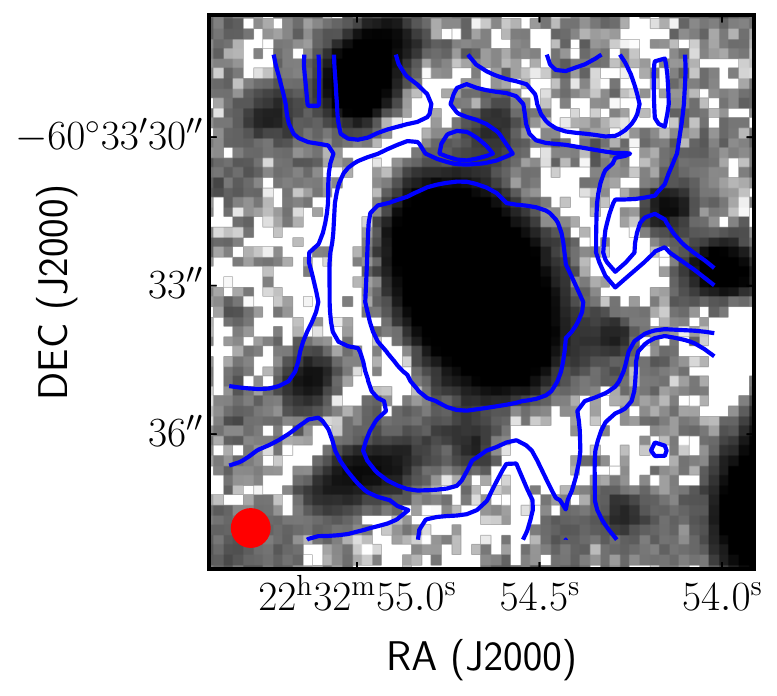}
\includegraphics[width = 0.32 \textwidth]{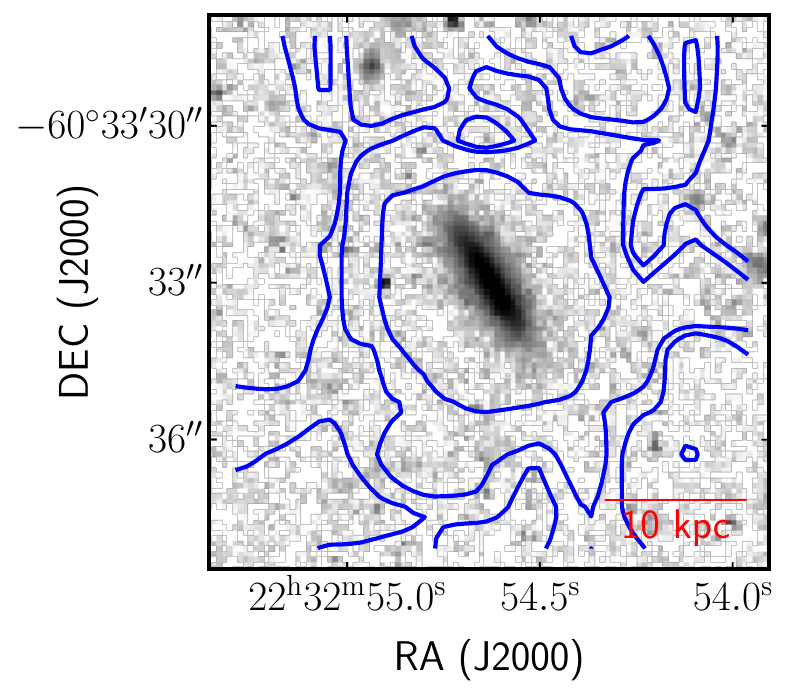}
\end{center}
\caption{{\bf The optical image of the MUSE HDFS obtained by integrating over the entire wavelength range (left), a zoom-in image of the target galaxy (middle) and the HST F606 image of the same zoom-in region (right)}.  
In the left panel, we also highlight the other two systems that will be discussed: two interacting galaxies at $z = 1.2840$ are marked with open circles and the Ly $\alpha$ emitter at $z = 3.9076$ with an open triangle. 
In the middle panel, which provides an enlarged view ($10^{\prime \prime} \times10^{\prime \prime}$, roughly $30\times30$ kpc$^2$) of our target, the image is logarithmicaly scaled and the three contours represent the H$\alpha$ emission at levels of 0.5, 2.5 and 12.5 in units of $10^{-19}$ erg\,cm$^{-2}$\,s$^{-1}$\,arcsec$^{-2}$. The FWHM of the PSF is represented by the red solid circle at the lower left. The right panel, which shows the HST image, demonstrates that our target is the dominant source in the region and that it is edge-on and undisturbed. The H$\alpha$ contours are reproduced for comparison, but we note that they are derived from the ground-based data and therefore are at lower angular resolution than the underlying image in this panel.} 
\label{fig:opticalMap}
\end{figure}

We highlight this galaxy and superimpose a blue contour that represents the detected H$\alpha$ emission at a level of $2.5 \times 10^{-19}$\,erg\,cm$^{-2}$\,s$^{-1}$\,arcsec$^{-2}$ in the left panel of Figure \ref{fig:opticalMap}.  
In the middle panel of Figure \ref{fig:opticalMap}, we present an expanded view ($10^{\prime \prime} \times10^{\prime \prime}$, roughly $30\times30$ kpc$^2$) of the optical image, in logarithmic scale to better resolve the low level optical emission, and additional contours of H$\alpha$ emission (0.5, 2.5 and 12.5 in unit of $10^{-19}$\,erg\,cm$^{-2}$\,s$^{-1}$\,arcsec$^{-2}$). H$\alpha$ emission is clearly detected and, pending additional discussion below, more extended than the broad-band (stellar) optical luminosity. Both are far broader than the point spread function (PSF), which is represented using a red circle with a radius of the full width at half maximum (FWHM) in the middle panel and discussed in further detail below. 
Finally, in the rightmost panel we present the F606W band HST image \cite{HDF-SWilliams, HST-HDFS}, which highlights the isolated nature of the galaxy. 

\begin{figure}[ht]
\begin{center}
\includegraphics[width = 0.48 \textwidth]{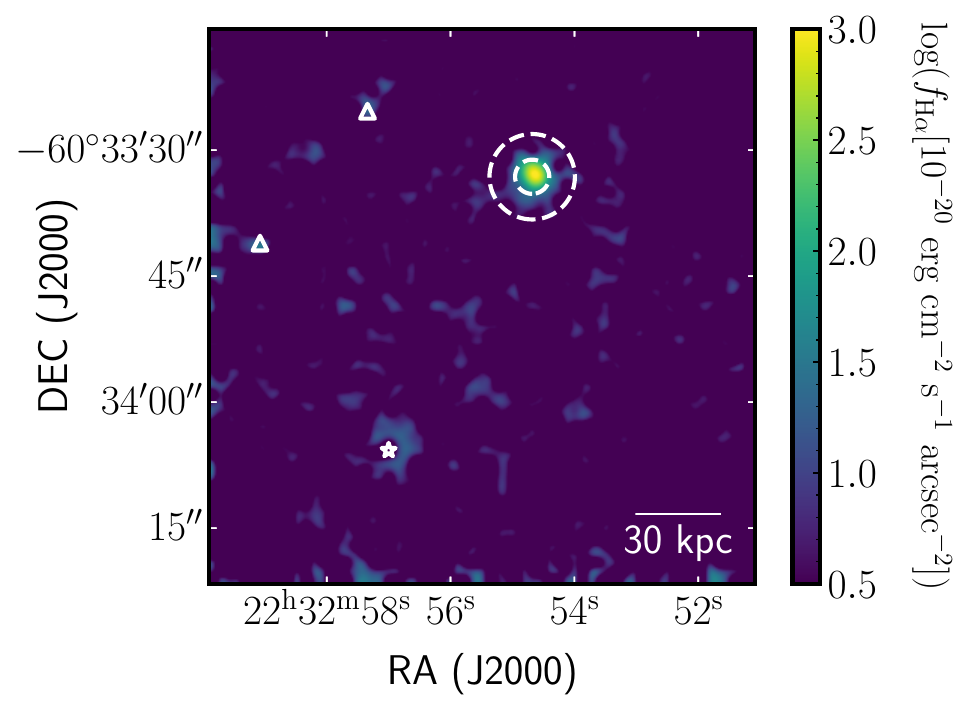}
\includegraphics[width = 0.48 \textwidth]{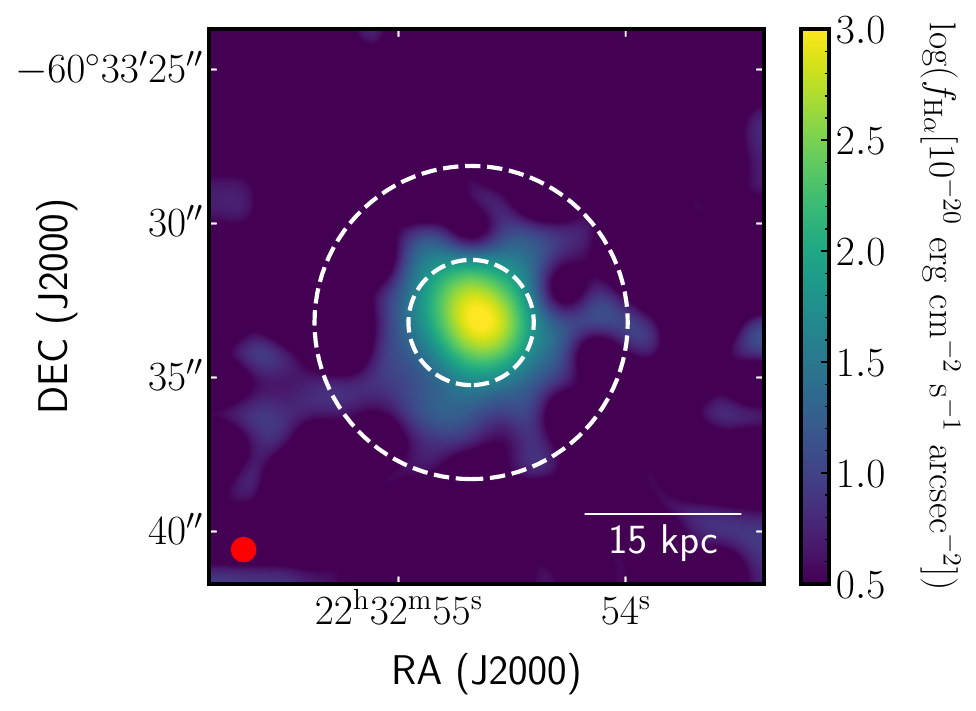}
\end{center}
\caption{{\bf The H$\alpha$ emission map of the HDFS field at the redshift of the target galaxy (left) and the zoom-in of the H$\alpha$ emission of the target galaxy (right)}. The two triangles mark the locations of the two low luminosity galaxies with similar redshift as the target galaxy and the star marks the location of the bright foreground star. The dashed circles have radii of 6 and 15 kpc and are centered on the location of the target galaxy. The galaxy disk size is $\lesssim$ 6 kpc \cite{Contini2016} and the H$\alpha$ emission can be significantly detected to 15 kpc as shown in Table \ref{tab:comparison}. The color mapping reflects a logarithmic scaling of the H$\alpha$ emission flux as shown in the sidebar.
The image plotting process does impose a bicubic interpolation that does slightly affect the visual impression but it is not relevant to our subsequent discussion regarding the extent of the emission. The FWHM of the PSF is represented by the red solid circle at the lower left.} 
\label{fig:HaMap}
\end{figure}

We present the image of the H$\alpha$ emission at the redshift of the target galaxy across the entire HDFS and the zoom-in on our low-redshift target galaxy in Figure \ref{fig:HaMap}. 
The target galaxy is the dominant source in Figure \ref{fig:HaMap} and the only emission feature that is clearly significantly above the background fluctuations.  We also highlight the locations of the two cataloged, lower luminosity galaxies with H$\alpha$ within the redshift window, although these do not correspond to positions of significant positive flux.
One is $\sim$ 50 kpc away from the target galaxy with $z = 0.1718$ and M606 = 25.44 [AB mag], another one is $\sim$ 100 kpc away from the target galaxy with $z = 0.1723$ and M606 = 25.72 [AB mag]. Neither of these galaxies satisfy the criteria in our stacking analysis. Quantitatively, the average H$\alpha$ emission surface brightness within a projected radius of 6 kpc from each of these two galaxies is $(3.1 \pm 1.6) \times 10^{-19}$ and $(5.2 \pm 2.7) \times 10^{-19}$\,erg\,cm$^{-2}$\,s$^{-1}$\,arcsec$^{-2}$, respectively. Both measurements are within 2$\sigma$ of zero flux. Finally, there is a slight residual structure at the location of the bright foreground star, presumably due to less than perfect continuum subtraction, but it is well below the residual flux level of the target even though it is a much brighter source in the optical image. There is no residual flux from the bright foreground star at the wavelengths of [O {\small III}] and [N {\small II}].

\begin{figure}[ht]
\begin{center}
\includegraphics[width = 0.46 \textwidth]{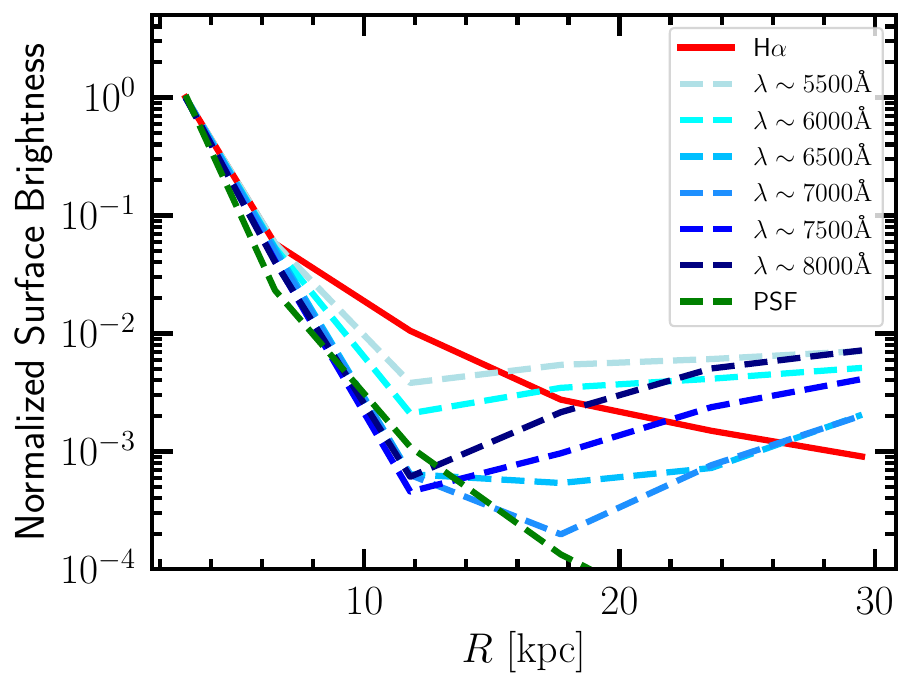}
\includegraphics[width = 0.48 \textwidth]{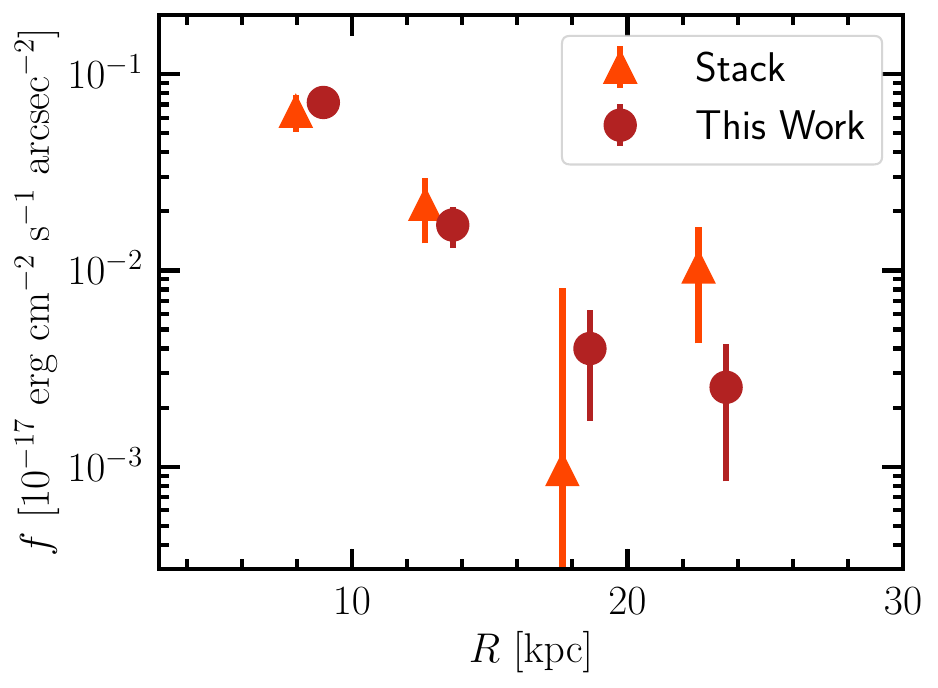}
\end{center}
\caption{{\bf Comparison of the radial profile of the H$\alpha$ emission around our low-redshift target galaxy to radial profiles of the same galaxy using six different continuum windows of width $\sim$ 600 \AA \ (left) and the comparison of H$\alpha$ emission flux as a function of projected radius between an SDSS spectral stack for low-mass galaxies and this work using MUSE spectra (right).}  The central wavelengths are shown in the legend. In the left panel, despite the higher background in the continuum profiles relative to that of the H$\alpha$ profile, as evidenced by their values at a radius of 30 kpc, they all lie below the H$\alpha$ profile for radii between 10 and 15 kpc, indicating that the extended H$\alpha$ emission is real and not a related to the stellar extent of the galaxy. }
\label{fig:radial}
\end{figure}

In the left panel of Figure \ref{fig:radial}, we present the radial profiles of the target in the H$\alpha$ image and of the target galaxy in a continuum image for different wavelength ranges, and the PSF for comparison. The PSF shape is circular and fitted using a Moffat function with $\beta$ parameter of 2.6. The FWHM is 0.66 arcsec at 7000 \AA, and  the average Gaussian white-light full FWHM value for the 54 exposures
is 0.77 $\pm$ 0.15 arcsec. The filter width in each continuum image is approximately 600 \AA. 
To calculate the profile of the target galaxy in both the H$\alpha$ and the optical image, we mask the nearby [O {\small II}] emitter  and other objects according to the HDFS catalog\cite{MUSE-HDFS} to mitigate contamination (more of an issue in determining the broad-band optical profile). Before discussing the comparison, we note that it is not straightforward. For the H$\alpha$ profile we did a continuum subtraction in each spectrum using 1000 \AA\ window surrounding H$\alpha$. As such, these spectra are always background subtracted and the radial profile should asymptote to 0 at large radius. On the other hand, the radial profiles of the continuum images have no such subtraction and they should therefore asymptote to the flux value of the unresolved background emission. For this reason, the optical continua radial profiles all lie above the H$\alpha$ profile at large radii. 

The H$\alpha$ signal lies above all of the broad-band optical profiles 
between projected radii of 10 and 15 kpc, establishing that the H$\alpha$ emission does not originate either from a central unresolved source or from sources distributed as the stars within the galaxy. Furthermore, the stellar continuum radial profiles are consistent with the PSF profile at small radii, demonstrating that our processing of the data recovers the expected radial profile for unresolved sources and that the stellar component of our target galaxy is at best slightly resolved. Determining whether the H$\alpha$ profile extends beyond 15 kpc requires a full understanding of the optical background to do the proper subtraction for the comparison profiles. If, however, we take the simple approach of subtracting a constant, such that all of the profiles agree at 30 kpc, then it is clear that we would conclude that the H$\alpha$ profile extends well beyond a radius of 15 kpc.

\begin{table}[ht]
    \centering
    \begin{tabular}{ccc}
    \hline \hline
    $r_p$ & Sample & $f_{\rm H \alpha}$ \\
    kpc &  & $10^{-19}$\,erg\,cm$^{-2}$\,s$^{-1}$\,arcsec$^{-2}$ \\  \hline
    \multirow{2}{*}{7.9} & Stack &  $6.47 \pm 1.30$ \\  
& MUSE & $7.17 \pm 1.13$  \\ \\
\multirow{2}{*}{12.7} & Stack & $2.17 \pm 0.78$  \\ 
& MUSE & $1.70 \pm 0.44$  \\ \\
\multirow{2}{*}{17.6} & Stack & $0.10 \pm 0.07$ \\
& MUSE & $0.40 \pm 0.23$  \\  \\
\multirow{2}{*}{22.6} & Stack & $1.04 \pm 0.59$ \\
& MUSE & $0.25 \pm 0.17$  \\ \hline
    \end{tabular}
    \caption{{\bf Comparison of the H$\alpha$ emission radial profile} between previous statistical work using SDSS spectra, for galaxies of comparable stellar mass, and this work across four independent radial annuli. }
    \label{tab:comparison}
\end{table}

To compare the H$\alpha$ radial profile to the results from the earlier stacked analyses of galaxies, we compare the mean H$\alpha$ emission flux within four annuli ($5.0 \le r <  10.0, 10.0 \le r < 15.0, 15.0 \le r < 20.0$ and $20.0 \le r < 25.0$ kpc) for this  galaxy to that corresponding to that of the `average' galaxy observed with SDSS
in Table \ref{tab:comparison} and the right panel of Figure \ref{fig:radial}. We mask a nearby [O {\small II}] emitter and other optical objects at  different redshifts 
according to the HDFS catalog\cite{MUSE-HDFS}  to mitigate possible contamination, for more details see the Methods section. The four annuli have mean projected radii of 8.2, 12.7, 17.6 and 22.6 kpc, respectively. 
For our SDSS comparison, we select galaxies with stellar masses in the range of $10^{8.0}$ and $10^{9.5}$ M$_\odot$ (mean stellar $= 10^{9.28}$ M$_\odot$)  to
create an SDSS stack for which the mean stellar mass is as close to  that of the MUSE target galaxy as possible.
Our MUSE results are statistically consistent  with the SDSS stack results (Table \ref{tab:comparison} and the right panel of  Figure \ref{fig:radial}), supporting an interpretation of the flux seen here at larger radii ($\gtrsim$15 kpc) as originating from the CGM. 
Furthermore, the agreement between the two radial profiles (the right panel of Figure \ref{fig:radial}) demonstrates that the limited physical resolution of the MUSE data for this particular galaxy (after binning, see Methods) is not impacting our measurement of the extended H$\alpha$ emission beyond $r \sim 6$ kpc.

\begin{figure}[ht]
\begin{center}
\includegraphics[width = 0.48 \textwidth]{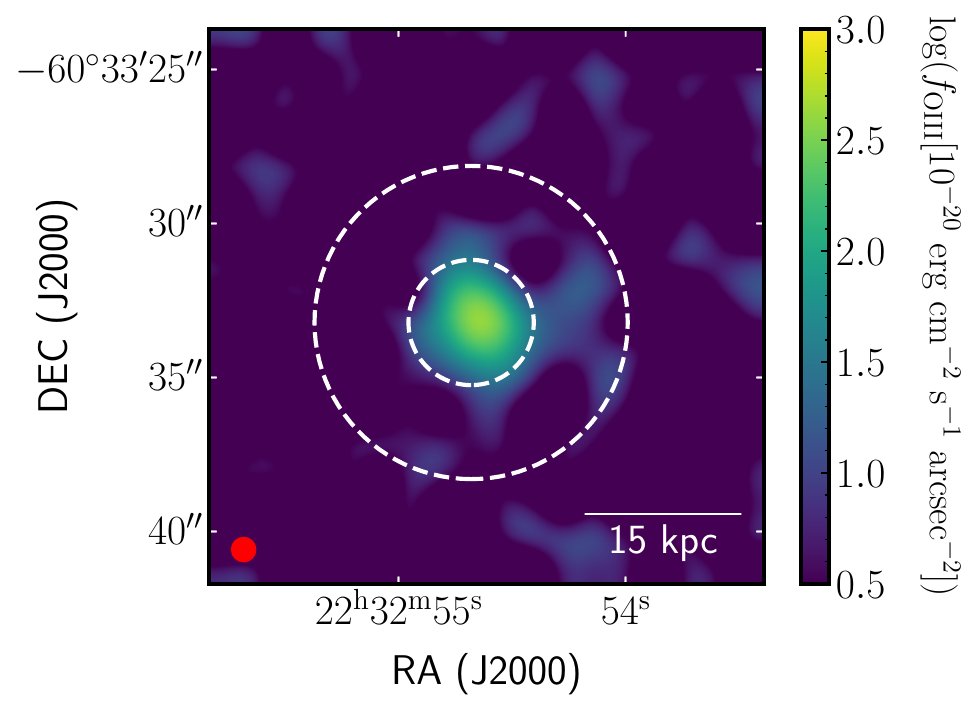}
\includegraphics[width = 0.48 \textwidth]{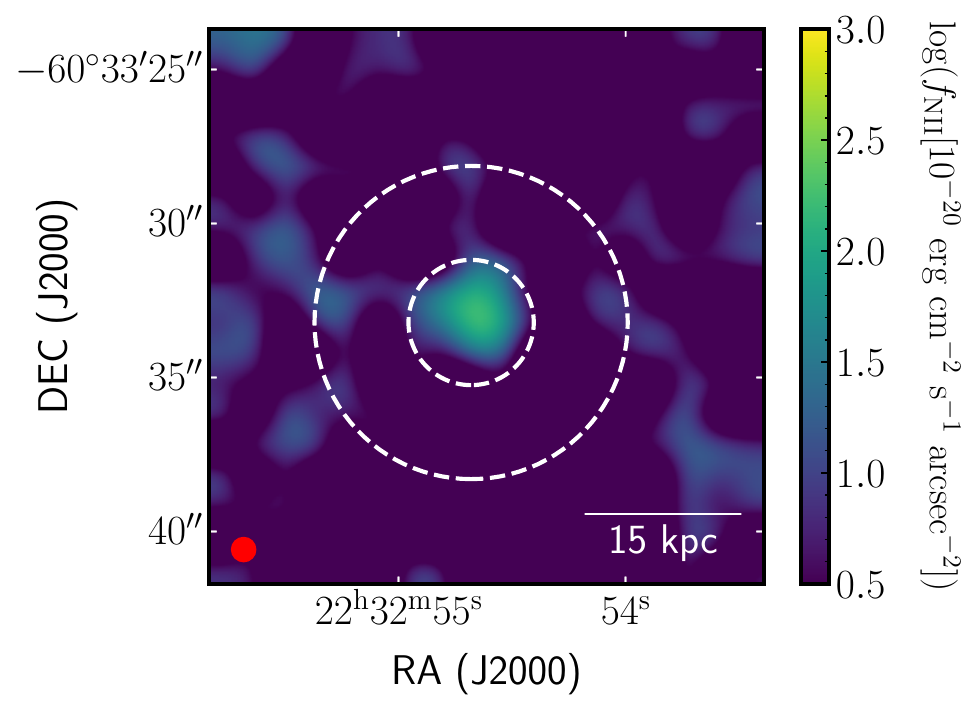}
\end{center}
\caption{{\bf The [O III]$\lambda$5007 (left) and [N II]$\lambda$6583 (right) emission images of our target galaxy}. The dashed circles correspond to projected radii of 6 and 15 kpc and are centered on the location of the target galaxy. The color mapping reflects a logarithmic scaling of the [O III]$\lambda$5007 and [N II]$\lambda$6583 emission flux, respectively,  as shown in the sidebar. The FWHM of the PSF is represented by the red solid circle at the lower left.} 
\label{fig:OIIIMap}
\end{figure}

Because the cool CGM is, in general, also expected to have emission lines other than H$\alpha$ emission, we  measure the residual fluxes at the locations of [O III]$\lambda$5007 and [N II]$\lambda$6583. However, as previous studies\cite{Zhang2018b} suggest, the CGM for a low-mass galaxy, such as this one, will be at best weakly detectable in [O III]$\lambda$5007 and not detectable in [N II]$\lambda$6583. Consistent with that caution, we find [O III]$\lambda$5007 emission only within the disk-CGM interface and no [N II]$\lambda$6583 emission flux on any scales greater than $\sim 5-6$ kpc (Figure \ref{fig:OIIIMap}).

\begin{figure}[ht]
\begin{center}
\includegraphics[width = 0.48 \textwidth]{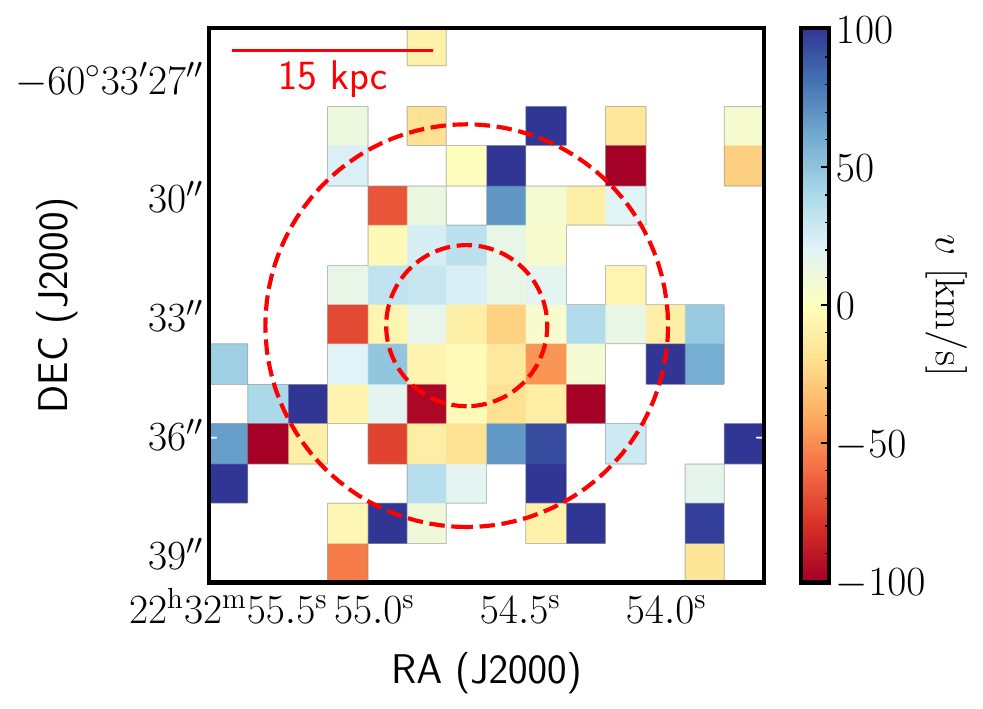}
\includegraphics[width = 0.47 \textwidth]{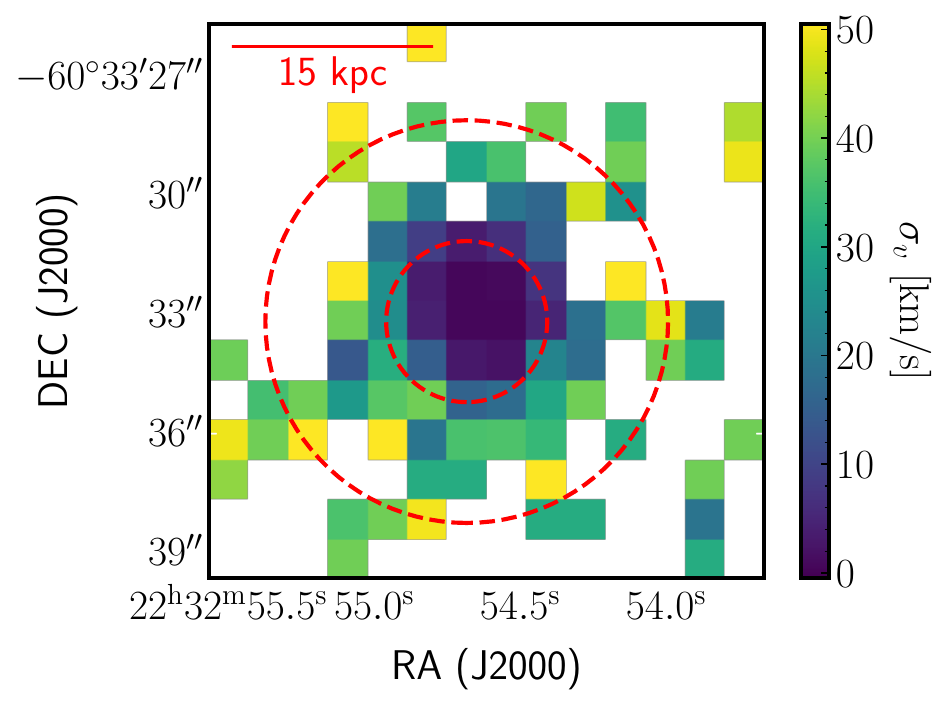}
\end{center}
\caption{{\bf The velocity field (left) and the uncertainty of the velocity field (right) of the target galaxy}. The dashed circles correspond to a projected radius of 6 and 15 kpc and are centered on the location of the target galaxy.}  
\label{fig:HaVel}
\end{figure}

Finally, we present the velocity field of this galaxy, as measured from the H$\alpha$ emission line in binned spaxels where it has a surface brightness greater than $0.5 \times 10^{-19}$\,erg\,cm$^{-2}$\,s$^{-1}$\,arcsec$^{-2}$, an area on the sky that roughly extends to our outer detection radius of 15 kpc, and the associated uncertainties in Figure \ref{fig:HaVel}. Our sensitivity limit is slightly lower than that quoted by Bacon et al. (2015) \cite{MUSE-HDFS} because we are using spaxels binned $5 \times 5$. The velocity within the galaxy disk (interior to the inner dashed circle) ranges between $\pm 40$ km s$^{-1}$, which is consistent with the results of Contini et al. (2016)\cite{Contini2016} and our classification of this galaxy as an LMC analog. At larger radii, the measured velocity range is larger, $\pm$ 100 km s$^{-1}$. Although the uncertainties are also larger at these radii, there are spaxels with these larger velocities for which $\sigma_v < 20$ km s$^{-1}$ suggesting the increase in velocity dispersion is not entirely a product of the larger uncertainties. We do not find clear evidence for coherent bulk motion, suggesting more chaotic or turbulent motion in the outer regions than in galaxy disk, although higher precision data are clearly needed to reach definitive conclusions. A path forward may be to forward model the velocity field of the CGM using hydrodynamic and magnetohydrodynamic (MHD) simulations such as those presented by \cite{Buie2022} and use the observations as a constraint on the physical models. Finally, we note that the gas velocities beyond a radius of 6 kpc do not resemble the disk rotation velocities, providing more evidence that the gas we detect beyond 6 kpc is not simply an extended disk  of star forming knots \cite{ferguson,thilker,z07,christlein}.

We interpret the emission at large radius as originating in the CGM of the galaxy for five reasons. First, the flux extends beyond the PSF profile and the galaxy itself has no extended stellar component, nearby companions, or obvious tidal features (Figure \ref{fig:opticalMap}. Second, H$\alpha$ emission is found at more than twice the optical radius of the galaxy. Third,  
the H$\alpha$ flux profile that is measured is in agreement with what that measured for a typical stacked SDSS galaxy of comparable stellar mass (the right panel of Figure \ref{fig:radial}, confirming that this is not an unusual galaxy. 
Fourth, the flux level is consistent with that expected from simple models of the CGM of such galaxies \cite{Zhang2021}. Finally, the kinematics of the gas beyond 6 kpc does not track the rotation of the inner disk.

Clearly a CGM extent as large or larger than $\sim$ 20 kpc, as we observe here, is not unreasonable for this galaxy.  The multi-phase CGM of the LMC, as probed with a sample of 28  {\sl Hubble Space Telescope Cosmic Origins Spectrograph (COS)} spectra of background UV-bright quasars\cite{Krishnarao2022} extends to a radius of at least 35 kpc. The multi-phase CGM of much larger galaxies, such as the Andromeda galaxy,  probed with a sample of 31 COS spectra of QSO sightlines\cite{Lehner2015} and  of 43 COS spectra QSO sightlines\cite{Lehner2020} extends to beyond the virial radius. More active galaxies tend to have easier to detect emission features. Using deep H$\alpha$ imaging from the Dragonfly Telephoto Array \cite{Dragonfly}, a giant shell of ionized gas at a projected distance of 40 kpc from the nucleus of the M82 was discovered \cite{Lokhorst2022}. Emission has also been detected, in the NGC 4631/4656 group \cite{Donahue1995}, in the starburst/merger NGC 6240 \cite{Yoshida2016}, a very low-mass (M$_* \sim$ $6\times 10^6$ M$_\odot$) blue compact dwarf galaxy \cite{Herenz2023} and a nearby star-burst galaxy \cite{Nielsen2023}. CGM emission is common in more active galaxies and likely to be common, at lower surface brightnesses, in normal galaxies.

An obvious question is whether the observed emission line fluxes plausibly correspond to emission from the CGM. To address this question we note that the observed radial H$\alpha$ profile (the right panel of Figure \ref{fig:radial}) matches what was observed in the SDSS stacks for similar galaxies and that \cite{Zhang2021} presented simple CGM models that reproduced both the observations and the gas cooling rates predicted by numerical simulations. As such, our interpretation of this emission arising in the CGM is viable.

\subsection{A 63 kpc wide Ly $\alpha$ feature and associated sources}
\label{sec:lya}

\begin{figure}[ht]
\begin{center}
\includegraphics[width = 0.48 \textwidth]{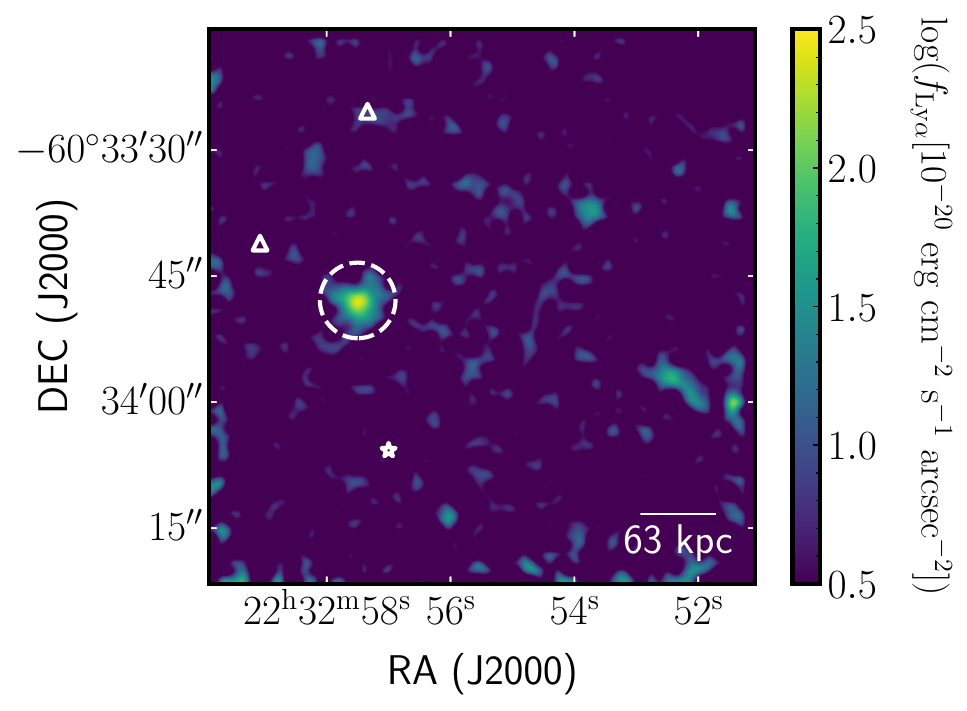}
\includegraphics[width = 0.48 \textwidth]{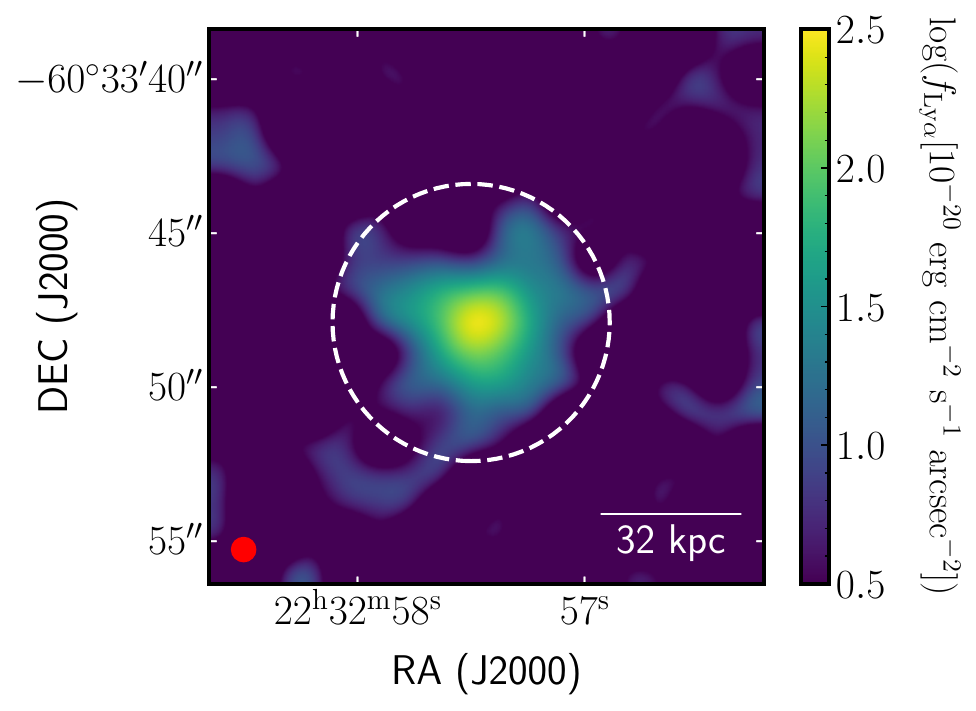}
\end{center}
\caption{{\bf The Ly $\alpha$ emission map of the HDFS field at the redshift ($z = 3.9076$) of the extended Ly$\alpha$ source (left) and the zoom-in (right) of the Ly $\alpha$ emission of the same source}. The Ly $\alpha$ source is located at $\alpha = 22 {\rm h} 32^\prime 57.5^{\prime \prime}$ and $\delta = -60^\circ 33^\prime 47.8^{\prime \prime}$ ($\alpha = 338.23962$, $\delta = -60.56346$). The circle corresponds to 9$^{\prime \prime}$ in diameter and 63 kpc at the redshift of the Ly $\alpha$ source. The FWHM of the PSF is represented by the red solid circle at the lower left.}  
\label{fig:LyaMap}
\end{figure}

We find a large (9 arcsec,  63 physical kpc wide)  Ly $\alpha$ emission structure associated with a source at $z=3.9076$\cite{MUSE-HDFS} at $\alpha = 22 {\rm h} 32^\prime 57.5^{\prime \prime}$ and $\delta = -60^\circ 33^\prime 47.8^{\prime \prime}$ ($\alpha = 338.23962$, $\delta = -60.56346$). 
The object is almost invisible in the optical image as shown in Figure \ref{fig:opticalMap}, 
and easily confused with a nearby brighter [O {\small II}] emitter at $z=0.4275$.  
The image of this source is presented in Figure \ref{fig:LyaMap}. 

This source mirrors the results of Guo et al. (2023)\cite{Guo2023}, who found evidence for large ($>$ 60 kpc) Ly $\alpha$ halos at redshift $3 < z < 4$ by stacking a sample of 155 Ly $\alpha$ emitters (LAEs) at redshift $3 < z < 4$ in the MUSE Extremely Deep Field (MXDF). Additionally, using extremely deep stacks of restframe-UV continuum and continuum-subtracted Ly$\alpha$ images for a sample of 92 UV continuum-selected, spectroscopically identified galaxies with $\langle z\rangle$ = 2.65, diffuse Ly$\alpha$ emission is detected to a radius of at least 80 kpc \cite{Steidel2011}. Our source is also reminiscent of the $\sim 100$ kpc Ly $\alpha$ structure, or ``blob", found in the MAMMOTH-1 field\cite{Cai2017} using 
the Keck Cosmic Web Imager (KCWI) on the 10-m Keck II telescope\cite{ZhangShiwu2023}
and the 97 kpc-wide H$\alpha$ structure emission found 
using the Multi-Object InfraRed Camera and Spectrograph (MOIRCS) on the 8-m Subaru telescope\cite{MOIRCS1,MOIRCS2}. A key difference between the published studies and ours is that the published ones targeted individual objects or structures where emission was expected, our study looked throughout the volume for emission. We did not find any that did not correspond to identified continuum objects down to our sensitivity, but not for lack of looking.
This use of IFUs complements narrow band searches\cite{Yang2004, yang09, Prescott2012} that are highly efficient in surveying across the sky for emission, but do not search across redshift. 

\subsection{A 130 kpc wide envelope surrounding a galaxy pair}
\label{sec:pair}

\begin{figure}[ht]
\begin{center}
\includegraphics[width = 0.48 \textwidth]{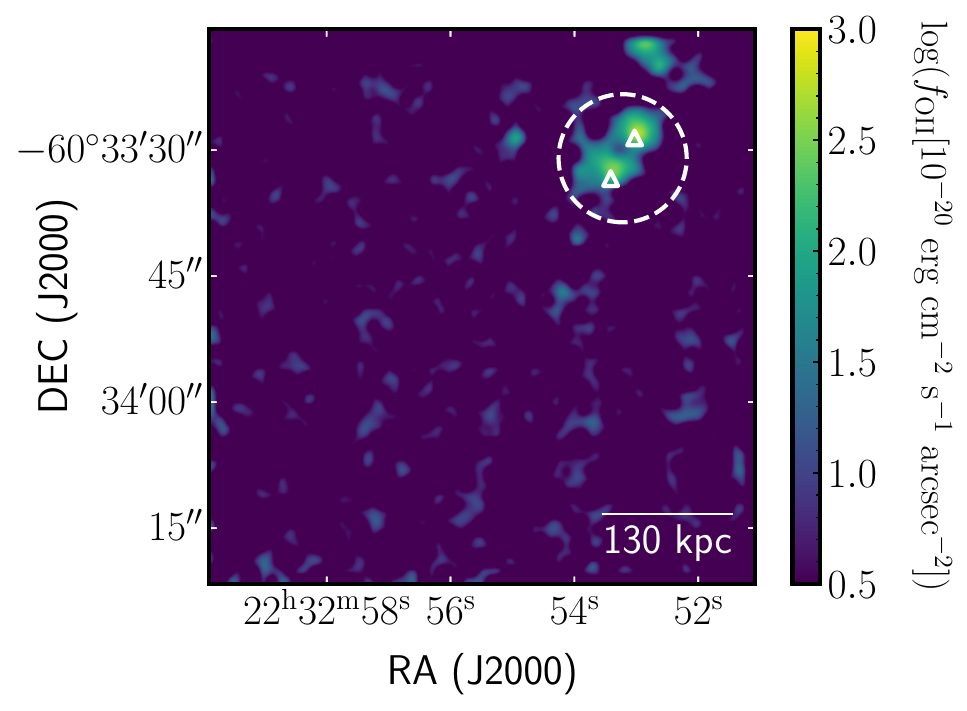}
\includegraphics[width = 0.48 \textwidth]{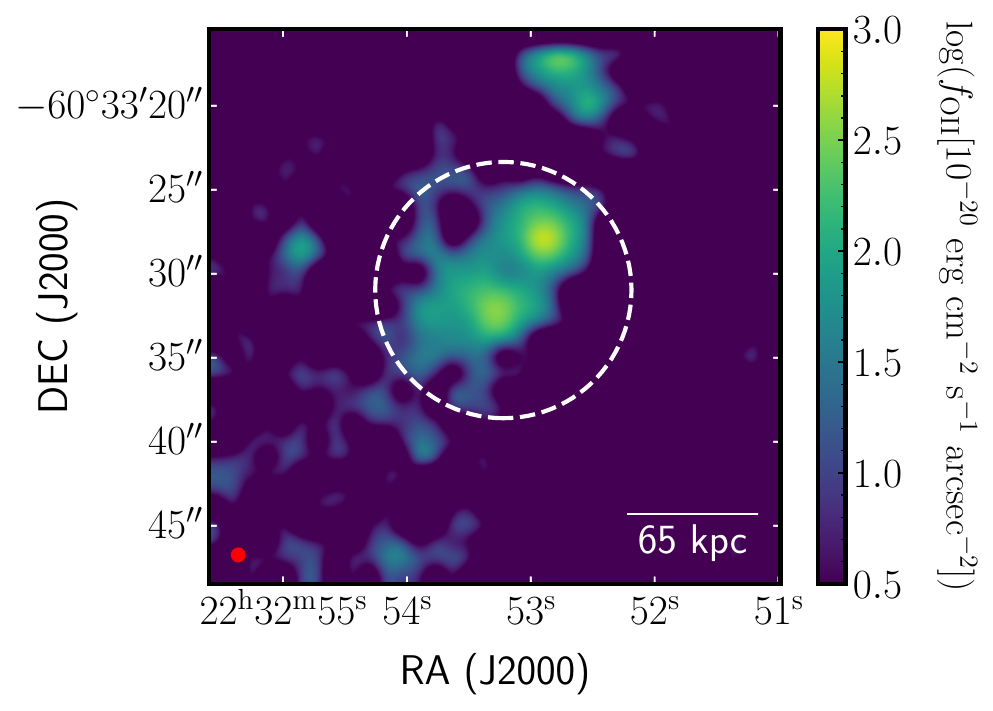}
\end{center}
\caption{{\bf The [O {\small II}]$\lambda \lambda$3727,3729 emission map of the HDFS field at the redshift ($z = 1.2840$) of a pair of interacting galaxies (left)  and the zoom-in (right) of the [O {\small II}]$\lambda \lambda$3727,3729 emission of the same source at the same redshift}. The circle corresponds to 15$^{\prime \prime}$ in diameter and 130 kpc at the redshift of the galaxy pair. The FWHM of the PSF is represented by the red solid circle at the lower left.}  
\label{fig:Galpair}
\end{figure}

We identify a pair of interacting galaxies at a redshift of 1.2840\cite{Contini2016} surrounded by a huge (15 arcsec, 130 physical kpc wide) envelope of [O II]$\lambda \lambda$3727,3729 emission 
(Figure \ref{fig:Galpair}). The stellar masses of the two galaxies are $10^{10.8}$ M$_\odot$ and $10^{10.3}$ M$_\odot$, and the SFRs are 2.0 M$_\odot$ yr$^{-1}$ and 11.5 M$_\odot$ yr$^{-1}$\cite{Contini2016}, respectively. Because the system encompasses two apparently interacting galaxies, we expanded the velocity window used to create the emission line map shown in Figure \ref{fig:Galpair} by two pixels ($\sim$ 100 km sec$^{-1}$) on either side of the mean [O II]$\lambda \lambda$3727,3729 recessional velocity of the galaxy pair. We note however that the full environment may be more complex because
there are three more galaxies at similar redshift (1.2838, 1.29, 1.2806) in the field.

For the smaller galaxy (lower left one) of the pair, we find an offset between the broad-band optical and the [O II]$\lambda \lambda$3727,3729 emission centers,  and that the smaller galaxy also has a longer extension of [O II] emission than the other, which might be an indication of the gas stripping due to the strong interactions between the two galaxies. In addition, there is emission seen to the north of the pair, near the upper edge of the image, that is associated with a galaxy at a similar redshift ($z = 1.2838$), suggesting an even more complex environment. We also find [C {\small II}]$\lambda 2326$ emission from this galaxy pair, suggesting that the interaction may be triggering emission from AGN activity.

\begin{figure}[ht]
\begin{center}
\includegraphics[width = 0.48 \textwidth]{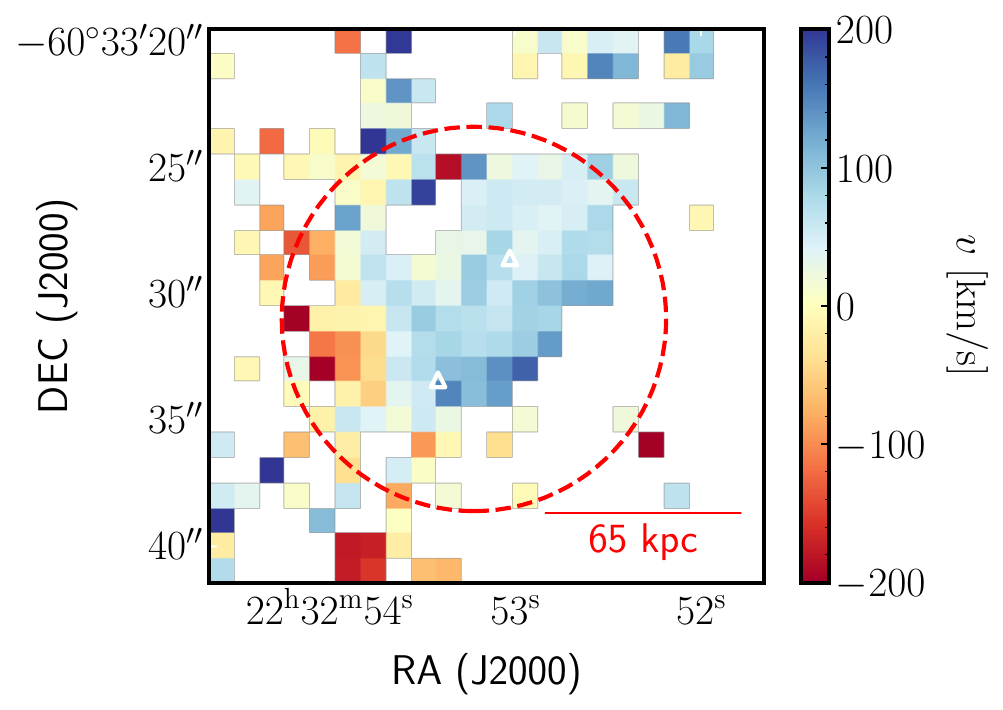}
\includegraphics[width = 0.48 \textwidth]{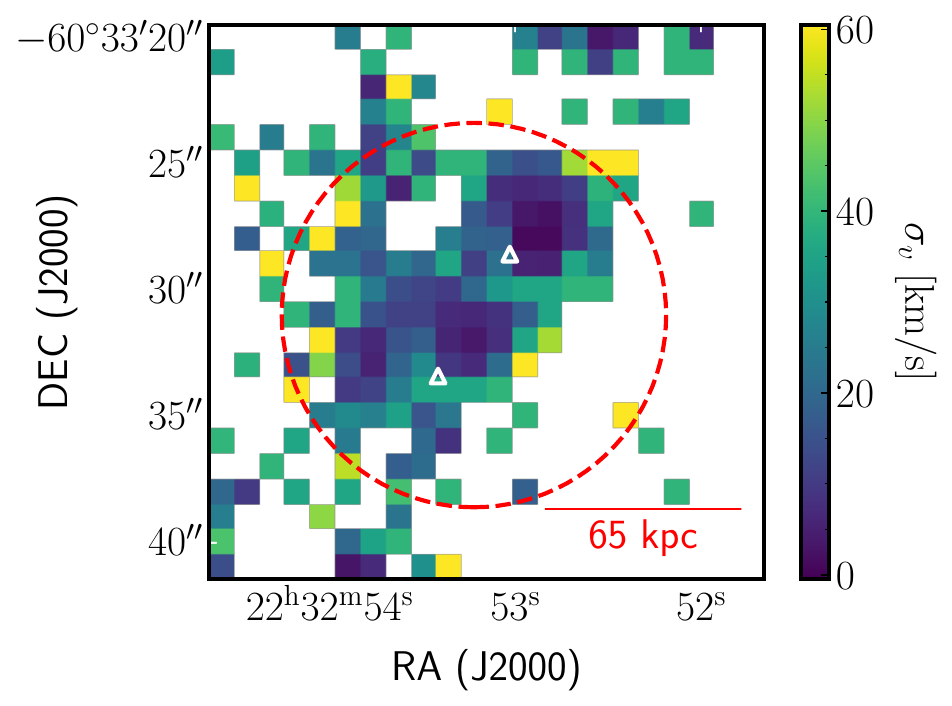}
\end{center}
\caption{{\bf The velocity structure (left) and  the uncertainty of the velocity field (right) of the interacting galaxies at the redshift of the galaxy pair}. The two triangles represent the position of the two galaxies in broad-band image. The circle corresponds to 15$^{\prime \prime}$ in diameter and 130 kpc at the redshift of the galaxy pair.}   
\label{fig:VelGalpair}
\end{figure}

In Figure \ref{fig:VelGalpair}, we show the velocity of the CGM surrounding the galaxy pair as estimated from the [O {\small II}]$\lambda \lambda$3727,3729 emission with flux greater than $0.5 \times 10^{-19}$\,erg\,cm$^{-2}$\,s$^{-1}$\,arcsec$^{-2}$. There is a suggestion of a velocity gradient, increasing velocity from east to west near the southern galaxy in the pair. The measured velocities, in the range of $\pm$ 150 km s$^{-1}$ from the mean, are consistent with what might be expected from the dynamics of an interacting system, but might also arise from hydrodynamic inflow or outflow of gas. The velocity difference between the two galaxies is $\sim$ 160 km/s, roughly consistent in magnitude with the velocity gradient we estimate, but the orientation of the gradient is not aligned with the galaxies. This misalignment suggests that the gas kinematics are complex, perhaps reflecting both tidal and hydrodynamical phenomena.

\section{Discussion}

These three examples highlight the variety of systems that can be uncovered in a source-blind survey for CGM emission. Furthermore, they begin to show the irregular structures and complex dynamics that are difficult to uncover unless one has contiguous data.
Although QSO absorption line work has been foundational in  revealing the IGM and CGM\cite{steidel2010,Chen2010,menard2011,bordoloi2011,zhu2013a,zhu2013b,Werk2013,Johnson2013,Johnson2014,Johnson2015,werk16,croft2016,prochaska2017,Cai2017,Chen2017a,Chen2017b,croft2018,lan2018,joshi2018,Chen2019,Zahedy2019,Dutta2020,Zheng2020,Haislmaier2021,CGMsquare2021,Norris2021,Qu2022}, the technique cannot provide contiguous 2-D maps or images of the CGM for normal, low redshift systems. 

Ours is not the first study to explore emission line imaging to trace the CGM, although previous work has targeted specific systems where one might have anticipated stronger than typical CGM emission. Examples include 
a galaxy group 
associated with a damped Ly $\alpha$ absorber\cite{Johnson2018, Chen2019} with possible gas stripping between the galaxy interactions, the region around the nearby starburst M 82\cite{Lokhorst2022}, a massive galaxy at $z=0.46$ with a star formation rate of 245 M$_\odot$ yr$^{-1}$ \cite{Rupke2019}, a star-burst sub-$L^*$ galaxy with a star formation rate of 12.1 M$_\odot$ yr$^{-1}$ \cite{Nielsen2023}, 
and a very low-mass blue compact dwarf galaxy\cite{Herenz2023} in the Local Universe. Where comparisons can be made, the results from these targeted studies are consistent with ours. For example, the blue compact dwarf, which could be a more active analog of our low mass, low redshift galaxy,  has emission identified out to 15 kpc\cite{Herenz2023} and the emission lines in two low mass, but interacting, galaxies in the intermediate redshift group also extend between 15 to 20 kpc\cite{Johnson2018, Chen2019}.

Although this work showcases the power of IFU spectroscopy to study the CGM of individual galaxies in detail, these MUSE data are far from the ideal  with which to probe the CGM of low redshift, normal galaxies with stellar mass of $\sim 10^{10}$ M$_\odot$. Exposures with similar time to the HDFS on  nearer and more massive galaxies (across 50 kpc or $0.25 r_{\rm vir}$ of the target galaxy) than that observed here will provide much more information than what can be gleaned from Figure \ref{fig:HaMap}, including the expected asymmetries and local fluctuations in the CGM\cite{corlies16}, the density profile, dynamics and kinematics, the ionization mechanism, chemical abundance patterns, and even possibly shocked regions arising from gas inflows/outflows\cite{Ford2014,Muratov2017,Nelson2019,Mitchell2020,Zhang2021, Herenz2023}. Long exposures ($\ge 25$ hours) with IFUs on 6- or 8-meter telescopes will reveal the CGM out to a radius of 50 kpc or $0.25 \ r_{\rm vir}$ and provide resolution in both radial and angular directions (a total of $\sim 30$ bins) for nearby (distances within a few tens of Mpc) galaxies. As we have mentioned, such work has already begun, although given the difficulty the targets tend to be starbursting or active galaxies (e.g., \cite{Burchett2021}). We hope that the results presented here, in particular those for the undistinguished galaxy that we presented in \S\ref{sec:low_mass}, help motivate the progression of such studies to ``normal", more representative galaxies.


In summary, we present a demonstration case for a source-blind observational approach in the study of the dominant baryonic component of galaxies. Although previous IFU observations have detected the CGM either in high-redshift\cite{Mackenzie2019,Kusakabe2022,Peroux2019} or lower-redshift extreme galaxies\cite{Yoshida2016}, we present this source-blind survey to probe a wide redshift range across the entire field for extended emission sources. Within the entire volume of HDFS we cataloged 63 emission structures, and we identified the CGM in an individual, low-redshift, normal galaxy (H$\alpha$ emission extending beyond a radius of 15 kpc), a large Ly $\alpha$ structure measuring over 63 kpc across at $z=3.9076$, and the 130 kpc wide CGM envelope surrounding two interacting galaxies at $z=1.2840$. We show that we are on the verge of using this type of observation to measure the nature of CGM velocities.  Long exposures with IFUs opens a new frontier in the exploration of the CGM even in representative galaxies and thus has tremendous promise for helping us unravel the mysteries of the baryonic cycle of galaxy evolution. 

\newpage         

\section*{Methods}
\label{sec:method}
The MUSE instrument provides 
high throughput (35\% end-to-end, including the telescope, at 7000 \AA),
moderate
spectral resolution (R $\simeq$ 3000 at $\sim$7000 \AA), full optical coverage
4650 $-$ 9300 \AA) spectroscopy
at a spatial sampling scale of 0.2 arcsec across a $1 \times 1$ arcmin$^2$ field-of-view\cite{MUSE2010}. The large field of view, wide wavelength coverage, and sufficient spectral resolution of the MUSE instrument make it well-suited for imaging the optical emission ([O {\small II}], [O {\small III}], H$\alpha$ and [N {\small II}]) originating from the cool CGM of individual lower-redshift, normal galaxies and [O {\small II}] or Ly$-\alpha$ emission from targets with much higher redshifts.

We use the public datacube obtained as part of the MUSE Hubble Deep Field South (HDFS) work\cite{MUSE-HDFS}. We briefly describe the observations and the data, for more details please see the original paper\cite{MUSE-HDFS}. The HDFS was observed  during six nights  with a total integration time of 27 hours covering the one arcmin$^2$ field of view centered at $\alpha = 22 {\rm h} 32^\prime 55.64^{\prime \prime}$ and $\delta = -60^\circ 33^\prime 47^{\prime \prime}$ ($\alpha = 338.23183$, $\delta = -60.56306$) (J2000). This location was selected to place one relatively bright star in the Slow Guiding System (SGS) and another star in the field of view to aid in the astrometric solutions. The spectrograph was rotated  90$^\circ$ after each integration, and the observations were dithered using random offsets within a 3 arcsecond box. The data were calibrated and reduced with version 0.90 of the MUSE standard pipeline\cite{Weilbacher2020}, including 9 steps that are elaborated in detail in Section 3.1 of the original paper\cite{MUSE-HDFS}. 

To increase the signal-to-noise sufficiently to detect the CGM, we bin the MUSE data along both the spectral and spatial dimensions. In the spectral direction, we extend the approach we used with the SDSS spectral stacks\cite{Zhang2016}. We measure the residual emission flux within an 8-pixel (10 \AA) sliding filter window in each spaxel
in the HDFS filed of view. The residual flux for each corresponding emission line at each specific redshift in each spaxel is determined by fitting and subtracting a third order polynomial to a 1000 \AA\ wide spectral section. As discussed in previous papers\cite{Zhang2016}, we require the continuum level for each spaxel spectrum to be less than 3 $\times$ $10^{-17}$ erg cm$^{-2}$ s$^{-1}$ \AA$^{-1}$ to limit the noise introduced by interloping strong emitters such as satellite galaxies or other interlopers. In the spatial direction we bin $5 \times 5 $ spaxels.


To identify sources in our effective narrow-band images, we use the commonly-used source extraction software SEP\cite{sep}, which is the Python version of SExtractor\cite{SExtractor}. We only catalog sources if they contain at least 5 contiguous pixels where each of those has a flux that is 5$\sigma$ above the mean background, leading to a source confidence that $> 5\sigma$. We remove spurious objects around the edge of the field by visual inspection.
Although the number of objects identified increases by almost a factor of two if we lower the detection threshold to 3$\sigma$, none of the additional objects are comparably extended to those that we have highlighted (i.e. $>$ 4 arcsec in radius). We present our catalog of the source-blind survey for emission structure in Table S1 in the Supplementary Materials. We match our catalog to the sources in the Bacon catalog\cite{MUSE-HDFS}, and finally we confirm, where possible, that the observed wavelength of the significant emission sources are consistent with the redshifts of their associated sources in the Bacon catalog\cite{MUSE-HDFS}. We define a radius for each emission structure by equating the area of a circular aperture of radius $r$ with the area identified by SEP to enclose the pixels that are  $> 2 \sigma$ flux above background. 

Next, we explore the possibility of false positives. We have focused our discussion on detected objects, but the vast majority of redshift slices contain nothing of significance, and so illustrate that false positives are rare. 
Figure \ref{fig:HaMap} provides a reasonably good guide of the magnitude of the noise in the H$\alpha$ map. However, we construct noise-only images by shifting the wavelength window slightly blue-ward or red-ward of the redshifted wavelength of H$\alpha$ for our low-redshift target galaxy in Figure S1 in the Supplementary Materials. There are no significant detections in either slice, including at the position of the low-redshift galaxies that we have discussed.

We now discuss two issues that are relevant to the determination of the radial emission profile of the low-redshift galaxy. First,
to estimate the uncertainties in the final emission flux values, we randomly select half of the spaxels in each radial bin, calculate the mean emission line flux, and repeat the process 1000 times to establish the distribution of measurements from which we quote the values corresponding to the 16.5 and 83.5 percentiles as the uncertainty range. We compensate for using only half the sample in each measurement by dividing the resulting $1\sigma$ estimated uncertainties by a factor of $\sqrt{2}$.
Second, we discuss possible contamination. There are two lower luminosity galaxies at a similar redshift which are about 4 magnitudes fainter than the target galaxy and are marked with  triangles in Figures \ref{fig:opticalMap} and \ref{fig:HaMap}. These are both projected sufficiently far from the target and have no emission of their own that they do not affect the measurements described here. 
Closer in projection is a relatively bright galaxy just to the lower right of our  galaxy of interest. It corresponds in location to a slight fluctuation in the H$\alpha$ image (Figure \ref{fig:HaMap}) but the galaxy is an [O II] emitter at a redshift of 0.5637\cite{MUSE-HDFS}. Given its redshift, there are no plausible emission lines that could be a contaminant but the source's presence could lead to higher noise at its location in the H$\alpha$ image. We therefore choose to mask it.
There are no indications of any other possible contaminating sources in the vicinity.

Finally, for the two objects in which we measure a velocity field (the low-mass galaxy at $z = 0.1723$ and the interacting galaxies at $z = 1.2840$), we adopt as the mean gas velocity in each spaxel the centroid of the fitted Gaussian to the emission line profile and its uncertainty. When the Gaussian fitting results in a velocity that is offset by more than 500 km s$^{-1}$ from the adopted mean velocity for the entire source (from \cite{MUSE-HDFS}) we consider the results to be suspect. In such cases, we adopt the flux-weighted centroid to estimate the mean gas velocity in that spaxel and use the flux errors to propagate the velocity uncertainty.

\bibliography{paper}
\bibliographystyle{Science}

\section*{Acknowledgments}
 HZ acknowledges financial support from the start-up funding of the Huazhong University of Science and Technology and the National Science Foundation of China grant (No. 12303007). DZ and HZ acknowledge financial support from NSF AST-2006785 and the University of Arizona. DZ and HZ acknowledge the MUSE team for the excellent observation and public data products. The authors thank the referees for comments that helped us improve the manuscript. The authors gratefully acknowledge the SDSS III team for providing a valuable resource to the community.
Funding for SDSS-III has been provided by the Alfred P. Sloan Foundation, the Participating I institutions, the National Science Foundation, and the U.S. Department of Energy Office of Science. The SDSS-III web site is http://www.sdss3.org/.

SDSS-III is managed by the Astrophysical Research Consortium for the Participating Institutions of the SDSS-III Collaboration including the University of Arizona, the Brazilian Participation Group, Brookhaven National Laboratory, Carnegie Mellon University, University of Florida, the French Participation Group, the German Participation Group, Harvard University, the Instituto de Astrofisica de Canarias, the Michigan State/Notre Dame/JINA Participation Group, Johns Hopkins University, Lawrence Berkeley National Laboratory, Max Planck Institute for Astrophysics, Max Planck Institute for Extraterrestrial Physics, New Mexico State University, New York University, Ohio State University, Pennsylvania State University, University of Portsmouth, Princeton University, the Spanish Participation Group, University of Tokyo, University of Utah, Vanderbilt University, University of Virginia, University of Washington, and Yale University.

\section*{Funding} 
HZ acknowledges  the start-up funding of the Huazhong University of Science and Technology and the National Science Foundation of China grant (No. 12303007). DZ and HZ acknowledge financial support from NSF AST-2006785 and the University of Arizona.

\section*{Author Contributions}
Both authors contributed to the final analysis and interpretation of the results. Huanian Zhang lead the data analysis. Dennis Zaritsky provided the initial motivation for the program. Both authors contributed equally.

\section*{Competing Interests}
The authors declare that they have no competing interests.

\section*{Data and Materials Availability}
 
All spectra data used in this paper are publicly available and the details are
presented in Methods and the original HDFS survey paper\cite{MUSE-HDFS} (Advanced data products are available at http://muse-vlt.eu/science). The original datacube were reduced using publicly available data reduction pipelines by the MUSE GTO team. The analysis code for the continuum fitting and image presentation is available.
Correspondence and requests for materials
should be addressed to Huanian Zhang ~(email: huanian@hust.edu.cn).

\clearpage

\begin{center}
    \bf \Huge Supplementary Materials for
\end{center}
\vspace{-0.75 cm}

\begin{center}
    \bf A MUSE Source-Blind Survey for Emission from the Circumgalactic Medium
\end{center}
\vspace{-0.75 cm}

\begin{center}
    Huanian Zhang$^*$ \& Dennis Zaritsky
\end{center}
\vspace{-0.75 cm}

\begin{center}
    $^*$Corresponding author. Email:huanian@hust.edu.cn
\end{center}

This PDF file includes:

Supplementary Text

1. Figure S1 to the Section of Method

2. Table S1 to the Section of Introduction and Method

The noise-only images by shifting the wavelength window slightly blue-ward or red-ward of the redshifted wavelength of H$\alpha$ for our low-redshift target galaxy is shown in Figure S1.

\begin{figure}[ht]
\begin{center}
\includegraphics[width = 0.48 \textwidth]{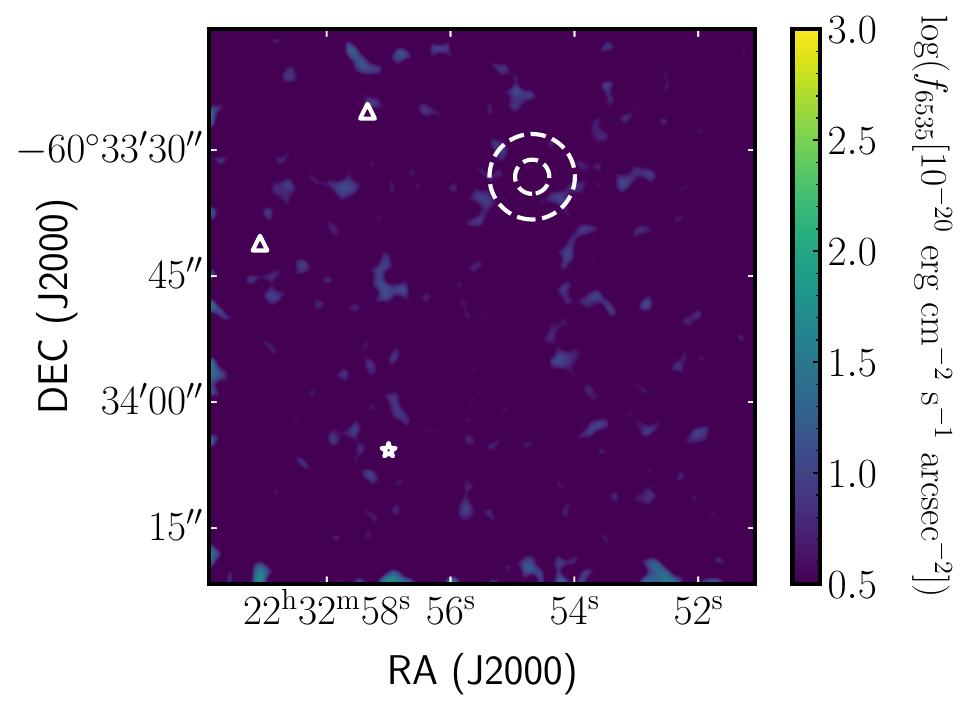}
\includegraphics[width = 0.48 \textwidth]{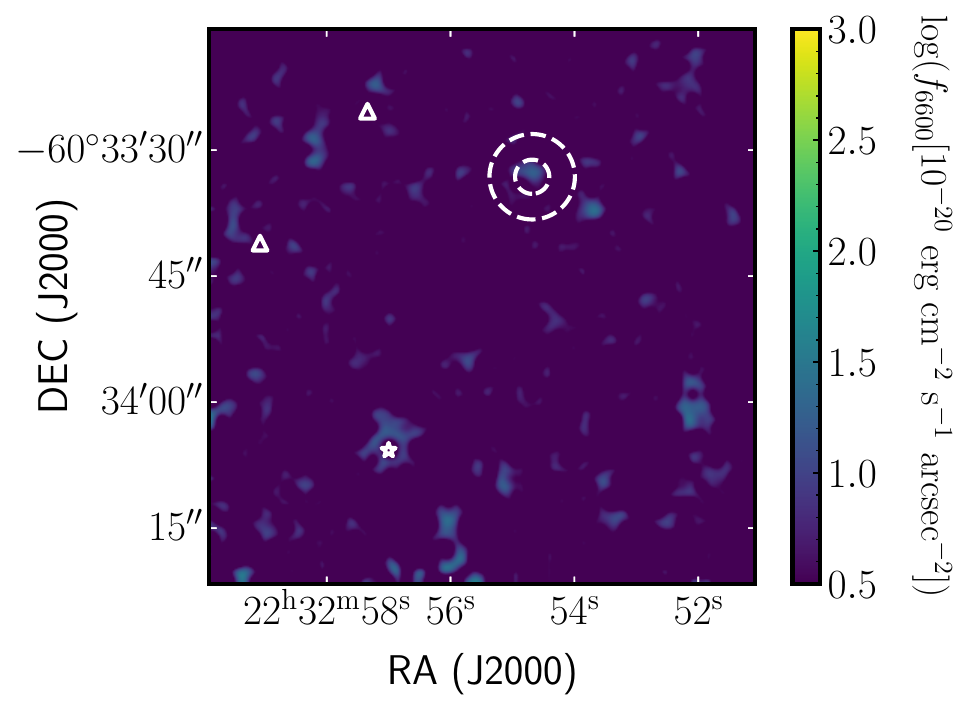}
\end{center}
\renewcommand\figurename{Figure S1}
\renewcommand{\thefigure}{}
\caption{ {\bf Narrow-band ``noise'' maps}. The left panel contains the map centered at 6535 \AA \ (approximately 30 \AA \ bluer of the H$\alpha$) and the right panel the map centered at 6600 \AA \ (approximately 40 \AA \ red-ward of the H$\alpha$). Note the level of the random fluctuations and the absence of any detection at our low-redshift source position.}
\label{fig:noise}
\end{figure}

The catalog (Table S1) consists of a total of 47 emission line sources, 43 of which are associated with previously identified continuum-detected sources and have measured redshifts.

\begin{table}[ht]
\begin{ThreePartTable}
\scriptsize
    \renewcommand\tablename{Table S1}
    \renewcommand{\thetable}{}
    \caption{{Catalog of Emission Line Detections}}
    \label{tab:catalog}
    \begin{tabular}
    {ccccccccccc}
    \hline\hline
    RA (J2000) & DEC (J2000) & $z$ & scale & Ly$\alpha$ & [O II] & H$\beta$ & [O III]$\lambda$4960 & [O III]$\lambda$5007 & H$\alpha$ & [S II]\\ 
    (degrees) & (degrees) & & (kpc/${\prime \prime}$) & ($^{\prime \prime}$) & ($^{\prime \prime}$) & ($^{\prime \prime}$) & ($^{\prime \prime}$) & ($^{\prime \prime}$) & ($^{\prime \prime}$) & ($^{\prime \prime}$)  \\  \\
    \hline\\
    338.21463 & -60.56043 & 0.57724 & 6.564 & - & 2.5 & 2.1\tnote{1} & - & 2.0 & - & - \\
    338.21518 & -60.56020 & 0.31778 & 4.633 & - & - & - & - & 2.3 & - & - \\
    338.21564 & -60.55840 & 3.47531\tnote{2} & 7.381 & 2.7 & - & - & - & - & - & - \\ 
     338.21680 & -60.56183 & 3.29034 & 7.480 & 3.0 & - & - & - & - & - & - \\
     338.21725 & -60.56654 & 0.56384 & 6.490 & - & 3.8 & - & 1.8 & 2.8 & - & - \\
     338.21735 & -60.55664 & 1.28997 & 8.370 & - & 3.0 & - & - & - & - & - \\ 
     338.21770 & -60.55541 & 0.31787 & 4.634 & - & 2.3  & - & -& 2.0 & 2.2 & - \\
     338.21793 & -60.56298 & 4.70057 & 6.510 & 2.3 & - & - & - & - & - & - \\
     338.21814 & -60.55918 & 0.57785 & 6.568 & - & 2.1 & 1.7 & - & 2.2 & - & - \\
     338.21902 & -60.56603 & 1.26581 & 8.352 & - & 2.0 & - & - & - & - & - \\
     338.21985\tnote{3} & -60.55576 &  &  & - & - & - & - & - & - & - \\
     {\bf 338.22095} & {\bf -60.55793}\tnote{4} & 1.28404 & 8.366 & - & 5.0 & - & - & - & - & - \\
     {\bf 338.22256} & {\bf -60.55927}\tnote{4} & 1.28527 & 8.363 & - & 5.5 & - & - & - & - & - \\
     338.22350 & -60.56528 & 3.08555 & 7.636 & 2.3 & - & - & - & - & - & - \\
     338.22390 & -60.56043 & 0.56375 & 6.490 & - & 4.6  & 2.5\tnote{1} & 2.6 & 4.2 & - & - \\
     338.22455 & -60.56722 & 0.99911 & 8.006 & - & 2.3 & -  & - & - & - & - \\
     338.22464 & -60.56286 & 0.32102 & 4.692 & - & 1.8 & -  & -& 2.0 & - & - \\
     338.22500 & -60.56152 & 1.28060 & 8.363 & - & 2.0 & -  & - & - & - & - \\
     338.22638 & -60.55575 & 0.56443 & 6.494 & - & 2.1 & - & -  & 2.0 & - & - \\
     {\bf 338.22784} & {\bf -60.55921}\tnote{5} & 0.17231 & 2.930 & - & -  & - & 2.1 & 2.6 & 4.8 & 1.5 \\
     338.22790 & -60.56514 & 4.69534 & 6.511 & 2.0 & - & - & - & - & - & - \\
     338.22840 & -60.57059 & 4.58043 & 6.550 & 2.5 & - & - & - & - & - & - \\

     338.22904 & -60.56020 & 1.15522 & 8.244 & - & 2.7 & - & - & - & - & - \\
     338.23007 & -60.56948 & 0.97225 & 7.953 & - & 2.8  & - & - & - & - & - \\
     338.23020 & -60.56875 & 0.46465 & 5.864 &  - & 2.9  & 2.5 & 2.6 & 2.6 & - & - \\
     338.23105 & -60.56117 & 3.34895 & 7.435 & 2.3 & - & -  & - & - & - & - \\
     338.23169 & -60.55536 & 4.31176 & 6.770 & 2.0 & - & -  & - & - & - & - \\
     338.23227 & -60.55937 & 0.56411 & 6.492 & - & 2.0 & -  & - & 2.3 & - & - \\
     338.23264 & -60.55839 & 0.22488 & 3.612 & - & - & - & -  & - & 2.6 & - \\
     338.23269 & -60.56849 & 3.12011 & 7.654 & 2.3 & - & -  & - & - & - & - \\
     338.23367 & -60.57060 & 0.56384 & 6.490 & - & 3.4  & 2.0\tnote{1} & - & 3.6 & - & - \\
     338.23523  & -60.55887 & 5.68513\tnote{2} & 5.879 & 2.0 & - & - & - & - & - & - \\
     338.23636 & -60.56071 & 4.01747 & 6.939 & 2.3 & - &  - & - & - & - & - \\ 
     338.23773 & -60.55639 & 0.67011 & 7.019 & - & 3.8  & 2.3 & - & 3.4 & - & - \\
     {\bf 338.23962} & {\bf -60.56346}\tnote{6} & 3.90764 & 7.017 & 4.5 & -  & - & - & - & - & - \\
     338.24194 & -60.56698 & 5.76414 & 5.836 & 2.6 & -  & - & - & - & - & - \\
     338.24255 & -60.55879 & 0.42232 & 5.551 & - & 3.2  & 2.3 & 2.7 & 3.4 & - & - \\ 
     338.24298 & -60.56303 & 1.21541 & 8.308 & - & 3.5  & - & - & - & - & - \\
     338.24448 & -60.55752 & 0.42227 & 5.553 & - & -  & - & - & 2.2 & - & - \\
     338.24124 & -60.56364 & 0.42754 & 5.591 & - & 2.9  & - & 2.1 & 2.9 & - & - \\
     338.24557 & -60.56668 & 4.01601 & 6.939 & 2.6 & -  & - & - & - & - & - \\
     338.24606 & -60.55707 & 3.33681 & 7.444 & 2.2 & - & - & - & - & - & - \\
     338.24610 & -60.57075 & {1.18472}\tnote{2} & 8.324 & - & 2.2 &  - & - & - & - & - \\ 
     338.24722 & -60.56849 & 3.27753 & 7.534 & 2.0 & -  & - & - & - & - & - \\      
     338.24762 & -60.56106 & 0.46367 & 5.857 & - & 3.5  & 2.8 & 2.8 & 3.1 & - & - \\
     338.24765 & -60.56558 & 1.02167  & 8.048 & - & 2.5 & -  & - & - & - & - \\
     338.24777 & -60.55804 & 1.09779  & 8.216 & - & 2.8 & -  & - & - & - & - \\
     \hline \hline
     \end{tabular}
     
\begin{tablenotes}\scriptsize
    \item[1] H$\gamma$ detection in addition to H$\beta$.
    \item[2] New redshift measurement.
    \item[3] The wavelength of the single-line emission structure is 8517.5 \AA.
    \item[4] Interacting galaxies at $z=1.284$ (see \S\ref{sec:pair}).
    \item[5] Low-mass galaxy at $z = 0.1723$ (see \S\ref{sec:low_mass}).
    \item[6] The extended Ly $\alpha$ emission at $z = 3.9076$ (see \S\ref{sec:lya}).
    \end{tablenotes}
    
\end{ThreePartTable}
\end{table}

\end{document}